\documentclass{aa}  
\usepackage{graphicx,psfig,amsmath}

\begin{document}
   \title{A diagram for the evaporation status of extrasolar planets}

   \author{A. Lecavelier des Etangs
           }
   \authorrunning{A. Lecavelier
           }

   \offprints{A.L. (\email{lecaveli@iap.fr})}

   \institute{Institut d'Astrophysique de Paris, CNRS, UMR7095 ; 
   Universit\'e Pierre et Marie Curie-Paris6, 98$^{\rm bis}$ boulevard Arago,
   F-75014 Paris, France} 

   \date{Received ...; accepted ...}

  \abstract
   {}
   {To describe the evaporation status of the extrasolar planets, we propose to consider
an energy diagram in which the potential energy of the planets is plotted versus the
energy received by the upper atmosphere.}
   {Here we present a basic method to estimate these quantities. 
For the potential energy, we include the modification of the gravity field 
by the tidal forces from the parent stars.}
   {This description allows a quick estimate of both the escape rate of the 
atmospheric gas and the lifetime of a planet against the evaporation process. 
In the energy diagram, we find an evaporation-forbidden region in which a gaseous planet 
would evaporate in less than 5 billion years. With their observed characteristics, 
all extrasolar planets are found outside this evaporation-forbidden region.
The escape rates are estimated to be in the range 
$10^{5}$\,g\,s$^{-1}$ to $10^{12}$\,g\,s$^{-1}$, 
with few cases above $10^{11}$\,g\,s$^{-1}$. 
The estimated escape rate for HD\,209458\,b is found to be consistent 
with the lower limit of $10^{10}$\,g\,s$^{-1}$ obtained from interpretation of
the H\,{\sc i} Lyman-$\alpha$ observations.

Finally, this diagram suggests possibilities for the nature of the recently 
discovered Neptune-mass planets. We find that GJ\,436\,b, 55\,Cnc\,e and 
HD\,69830\,b cannot be low mass gaseous planets. 
With density necessarily above 0.5\,g\,cm$^{-3}$ to survive evaporation,
these planets must contain 
a large fraction of solid/liquid material. Concerning GJ\,876\,d, we find that
it must have a density larger than $\sim$3\,g\,cm$^{-3}$ 
to survive the strong EUV energy flux from its nearby parent 
star. GJ\,876\,d must contain a large fraction of massive elements. 
}
   {}

   \keywords{Stars: planetary systems}


   \maketitle
%

\section{Introduction}
\label{Introduction}

Among the nearby two hundreds extrasolar planets identified today, there is a particular class 
of planets very close to their parent stars with semi major axis $\la$0.1\,AU. Among them, 
there are massive hot-Jupiters (also named ``Pegasides'', Guillot et al.\ 1996) and small mass 
planets (see for instance, Santos et al.\ 2004; Rivera et al.\ 2005). These planets constitute 
about 20\% and 5\% respectively, of the planets identified today. 
Since the discovery of the first of them, 51\,Peg\,b 
(Mayor \& Queloz 1995), the issue of their evaporation status has been raised (Burrows \& 
Lunine 1995; Guillot et al.\ 1996). Since these planets exist and
orbit stars which are not particularly young, it is clear that the evaporation rate of 
the observed massive
hot-Jupiters should be modest enough to not dramatically impact 
their evolution. At least they can survive billion years at the place where they
are observed.
It is extremely unlikely that we are witnessing a transient phase in the evolution of 
extrasolar planets (Baraffe et al.\ 2004). 
In brief, the discoveries of a large number of massive 
hot-Jupiters\footnote{usually defined as massive planets orbiting 
in less than 10~days} and 
very-hot-Jupiters\footnote{usually defined as planets orbiting in less than 3~days 
({\it e.g.},  Gaudi et al.\ 2005)}
led to the conclusion that the evaporation {of massive planets} has to be modest.

In that frame the discovery that the transiting extrasolar planet HD\,209458\,b is indeed losing 
mass was mostly unexpected (Vidal-Madjar et al.\ 2003). From the observation of 15$\pm$4\% 
absorption in Lyman-$\alpha$ during the transit, a lower limit of the H\,{\sc i} escape rate was 
estimated to be of the order of $\sim10^{10}$\,g\,s$^{-1}$ (Vidal-Madjar et al.\ 2003; 
Vidal-Madjar \& Lecavelier des Etangs 2004). This mass loss rate corresponds to a loss of 
$\sim$0.1\% of the total mass of the planet over 10 billion years (10 Gyr). Models of the time 
variation of this escape rate lead to slightly larger values for the total fraction of mass loss 
over the whole stellar life (see e.g., Lecavelier des Etangs et al.\ 2004; Baraffe et al.\ 2004).

Different detailed models have been developed to understand the observed evaporation 
process and to 
estimate the escape rate (e.g., Yelle 2004; Tian et al. 2005,  Garc\'\i a Mu\~noz 2006), 
to be compared with the 
observational constrains (Vidal-Madjar et al.\ 2003, 2004). Finally, it has been suggested that 
evaporation can lead to significant modification of the planets nature in the case of planets
with a mass only a fraction of the mass of the observed hot-Jupiter
and orbiting close to their parent stars (Lecavelier des Etangs et al.\ 2004; Baraffe et al.\ 
2005).

However, models were applied to specific cases; it is not yet
possible to conclude on the order of magnitude
of the possible evaporation of planets 
which are being discovered without the effort to model 
in detail each particular case.
The purpose of the present paper 
is not to review previous works and extended modeling efforts.
Here we propose an alternative approach to obtain general
characteristics from basic observed parameters. This approach
allow statistical conclusion without the need to model in detail
every individual planet for which detailed physical characteristics are not
available.   
We propose to outline the main characteristics of 
a given planet that allow concluding on the general trend of 
its evaporation status, 
escape rate and lifetime related to the erosion through evaporation.

Here we present the idea of an energy diagram to address these issues 
(Sect.~\ref{Potential energy of planets versus stellar energy flux}), 
and the way to simply estimate the energetic characteristics of a given
planet (Sect.~\ref{The extreme ultraviolet illumination from the central stars} 
and~\ref{Potential energy}). Then an evaluation of the
lifetime and escape rates is presented in Sect.~\ref{Lifetime and escape rate}.
The resulting diagram is shown in Sect.~\ref{The energy diagram}.
A diagram corresponding to the escape rate is given in 
Sect.~\ref{Diagram of the potential energy of the atmosphere and the corresponding escape rate}. 
Uncertainties and evolutionary tracks are presented
in Sect.~\ref{Uncertainties} and~\ref{Evolution tracks}. 
Finally we conclude on the possible nature of the
recently discovered Neptune mass planets (Sect.~\ref{Neptune mass planets in the diagram}).

\section{Potential energy of planets versus stellar energy flux}
\label{Potential energy of planets versus stellar energy flux}

A plot of the lifetime of planets as function of their masses and orbital distances has been 
calculated by Lecavelier des Etangs et al.\ (2004). The purpose of this plot is to show that the 
identified planets are stable against evaporation. For the existing planets, the lifetime, 
defined as the time to escape the full mass of the planet, is always longer than 
$\sim10^{10}$ years (Fig.~5 of Lecavelier des Etangs et al.\ 2004). 
Attempts have been made to extrapolate from this plot possible 
escape rate in the case of planets discovered later (HD\,149026\,b, Valenti 2005). It would 
be extremely valuable to have a quick diagnostics on the evaporation rate of a given planet to 
predict  
if there are any chances to observe and constrain the evaporation through, e.g.,  
Lyman-$\alpha$ observation during a transit (Lecavelier des Etangs et al., in preparation). 
However, in the lifetime plot of Lecavelier des Etangs 
et al.\ (2004), the energy flux from the star is assumed to be the one of HD\,209458\,b. 
The energy flux from other stars is certainly different depending on the stellar type, rotation 
period, etc. The second critical parameter in this plot is the assumed 
planetary radii. In the lifetime plot of Lecavelier des Etangs et al.\ (2004), the radius of 
HD\,209458\,b (1.35 Jupiter radii) 
was used to normalize the scaling laws for the planetary radii because 
at the time of this work HD\,209458\,b was the only hot-Jupiter whose radius  
had been measured. However, the radius of HD\,209258\,b is now considered as anomalously
large (Guillot 2005; Lecavelier des Etangs \& Vidal-Madjar 2006). From observations 
of ten
transiting planets (Bouchy et al.\ 2005b, McCullogh et al.\ 2006) and detailed modeling of 
the planetary structure (Guillot 2005), 1.1~Jupiter radii is certainly more appropriate for a 
1~Jupiter mass planet close to its parent star. 
This makes HD\,209458\,b with its 1.35 Jupiter radii and 0.69 Jupiter mass
clearly a peculiar object.
In the following we will therefore use
a fit to the Guillot's model. A new plot has been calculated with this new 
information (Lecavelier des Etangs \& Vidal-Madjar 2006). 
However, the planetary radius depends also on parameters other than the planetary mass and 
orbital distance, for instance, the stellar type 
and the corresponding effective temperature 
of the central star and its age. It is thus impossible to uniquely estimate lifetime and 
escape rate from the position of a given planet in the orbital distance versus mass diagram. 
As a consequence, this plot has to be revisited. 

In summary, since each planet has different radius and different energy flux 
entering onto the upper atmosphere from stars of different type, 
the orbital distance versus mass diagram is not appropriate 
to conclude on the escape rate and life time of a variety of planets. 
We have to define the parameters (or combination of parameters) which allow us 
to conclude on that matter.

After the realization that HD\,209458\,b is losing mass, many modeling efforts have been 
pursued to better understand the escape in the particular case of hot-Jupiters.  
The common results of these numerous models is that almost 100\% of the 
EUV/Lyman-$\alpha$ energy flux which is converted into heat 
is then transferred into escape (see, for 
instance, Yelle 2004 and Tian et al.\ 2005). 
Even magnetic fields which can inhibit ion escape are found to have modest impact.
Yelle (2004) found that the net escape rate with ion escape inhibited by magnetic field 
is only 30\% 
smaller than in its reference model.
Similarly, effects of dissociation and ionization of H$_2$ are found to be negligible.
The resulting escape rates of these various models 
are different not because these models assumed different mechanisms but because of the 
different illumination, planetary cross section for the energy flux, and 
fraction of this energy converted into heat to the upper atmosphere.

As a result, the evaporation status of a set of different planets orbiting 
different stars can be obtained by a comparison of the energy deposited in the upper atmosphere
and used to escape the planet gravity, {\it i.e.} to compensate for the (negative) potential 
energy of the atmospheric gas, with this potential energy.
We therefore propose to calculate the position of the extrasolar planets in a diagram 
displaying the potential energy versus the stellar energy deposition.
We now evaluate these two quantities in the two next sections to
plot the corresponding diagrams in Sect.~\ref{The energy diagram} 
and~\ref{Diagram of the potential energy of the atmosphere and the corresponding escape rate}.

\section{The extreme ultraviolet illumination from the central stars}
\label{The extreme ultraviolet illumination from the central stars}

The present and past illumination from the parent stars is a key 
parameter which needs to be accounted for. 
First, as the planet-hosting stars are cooler than early F stars in the main sequence, 
the stellar energy emitted in the near UV range is negligible (e.g., Kurucz 1993). 
Moreover, near UV and optical energy is absorbed deep in the bottom and dense atmosphere. 
On the contrary, neutral components of atmospheres have large cross section in the 
extreme ultraviolet (100\AA$<\lambda<$1200\AA).
Photons in the extreme ultraviolet wavelength range (EUV) 
are thus easily absorbed by the tenuous upper atmospheres of planets.
For instance, in the terrestrial planets of the Solar system, 
the EUV energy flux is the main driver of the escape (e.g., Chamberlain \& Hunten 1987).
Therefore, for simplicity and in agreement with the result of all
detailed models of the evaporation of hot-Jupiters, we can assume that 
the energy flux deposited in the upper atmosphere 
and transferred into escape energy mainly comes from the EUV.

However, because of the interstellar absorption, there are very few observational 
constraints on the EUV flux from the stars in the solar neighborhood. 
On the other hand, although stellar EUV fluxes are related 
to stellar rotational velocities (Wood et al.\ 1994), 
most of the stars harboring extrasolar 
planets have no published rotational velocities or only upper limits.
In most cases, the information on the planet-hosting stars is limited to the
stellar type. We have hence chosen 
to consider the very extreme ultraviolet luminosities 
$L_{S2}$ as a function of the stellar type, 
as measured by {\sl Rosat} in the S2 channel between 110 and 200~\AA .
We then scale these luminosities to the solar EUV flux of 
$\sim$4.6\,erg\,cm$^{-2}$\,s$^{-1}$ at 1~AU as given by Ribas et al.\ (2005).
This solar value is obtained considering mid-solar cycle data of 1993 
as representative of the Sun at the average activity.
The EUV flux, $F_{\rm EUV}(1\,{\rm AU})$, received by a planet by unit 
area at 1~AU is hence calculated using
\begin{equation}
F_{\rm EUV}(1\,{\rm AU})=4.6 \frac{L_{S2}}{L_{S2, {\rm G\,stars}}}\ {\rm erg}\,{\rm cm}^{-2}\,{\rm s}^{-1}.
\end{equation}
The scale between $L_{S2}$ luminosities and EUV is justified by the fact that both the 
$\sim100-200$\,\AA\ and the extreme-far UV fluxes are emitted by the same region in the 
chromosphere and the transition region where the temperature is observed to increase from 
the photospheric to the coronal values.

Hodgkin and Pye (1994) give median S2 luminosities as a 
function of the spectral types, with value of $10^{27.9}$, $10^{26.8}$, 
$10^{27.3}$ and $10^{26.6}$\,erg\,s$^{-1}$ for F, G, K and M stars respectively.
Concerning the F type stars, most of the F stars known to harbor planets are 
late type F8{\sc v} or F9{\sc v} stars. Because early F stars remain 
chromospherically active and fast-rotating during their main sequence phase, 
the accuracy in the measurement of their radial velocity is strongly 
limited. It is yet impossible to detect planets
around them (F.~Pont, private communication).
But these early F stars contribute 
to the mean value of the observed S2 luminosities of the whole class of F stars. 
We therefore decided to consider a typical S2 luminosity of $10^{27.3}$\,erg\,s$^{-1}$ 
for F6 and F7 stars harboring planets, and an S2 luminosity of $10^{26.8}$\,erg\,s$^{-1}$ 
for later F8{\sc v} and F9{\sc v} stars which are more similar to G0{\sc v} type stars.

Using these S2 luminosities and the solar EUV luminosity,
we find 
$F_{\rm EUV}(1\,{\rm AU})$=14.7, 4.6, 4.6, 14.7 and 2.9\,erg\,cm$^{-2}$\,s$^{-1}$ for 
F6-F7, F8-F9, G, K, and M stars, respectively.
Except if explicitly mentioned (Sect.\ref{Uncertainties}), these values will be used 
throughout the paper.
These fluxes can be converted to the energy per unit of time, $dE_{\rm EUV}/dt$, 
received by a planet of radius $R_p$ with an orbital distance $a_p$ using:
\begin{equation}
dE_{\rm EUV}/dt=\pi R_p^2 \left(\frac{a_p}{1\,{\rm AU}}\right)^{-2}F_{\rm EUV}(1\,{\rm AU}).
\label{E=E(F)}
\end{equation}

\section{Potential energy}
\label{Potential energy}

To evaluate the total lifetime of a planet in an energetic point of view, we need to 
consider the total potential energy, $E_p$, of the planet considered as a whole,
from its surface to its deep interior.
If this energy is given is {\it erg}, the ratio with the received energy 
expressed in {\it erg per billion of years} will provide an estimate of the
total time needed to cancel the (negative) potential energy, in other words,
the evaporation lifetime. On the other side, to evaluate the escape of the atmospheric
gas at the present time, we need to consider the potential energy
of the gas only in the atmosphere and per unit of mass. The ratio of this quantity
with the energy received will provide the mass escape rate.
We see that we have two different quantities which need to be estimated:
the potential energy of the whole planet and 
the potential energy per unit of mass in the atmosphere

\subsection{The potential energy of the whole planet}
\label{The potential energy of the whole planet}

The potential energy of a gaseous sphere of radius $R_p$ is given by 
\begin{equation}
E_p=-G\int_0^{R_p}\frac{m(r)}{r}\frac{d  m(r)}{d  r}dr
\end{equation}
where $G$ is the gravitational constant and $m(r)$ is the mass inside a radius $r$.
For a uniform density and a mass $M_p$, we obtain the classical formula 
$E_p=-3 GM_p^2/5R_p$. 
But in general, the potential energy depends on the internal structure.
For a planet with a mass $M_p$ and a radius $R_p$,
an analytical approximation can be found assuming a polytropic approximation
of the equation of state for the planet interior. Using a polytropic index $n=1$ 
($P\propto\rho^2$, de Pater \& Lissauer 2001) the potential energy is found to be:
\begin{equation}
E_p=-\frac{3}{4}G M_p^2/R_p.
\label{E_p}
\end{equation}
The details of the calculation are given in Appendix~\ref{App:Ep}.
Numerically this converts into:
\begin{equation}
E_p=-2.9\times 10^{43} 
 \left(\frac{M_p}{M_{\rm Jup}}\right)^2 \left(\frac{R_p}{R_{\rm Jup}}\right)^{-1} 
{\rm \ erg}.
\end{equation}

\subsection{The potential energy per unit of mass in the atmosphere}
\label{The potential energy per unit of mass in the atmosphere}

Another interesting quantity is the potential energy of the atmospheric gas per unit of mass. 
This potential energy per unit of mass is simply given by
(see Appendix~\ref{App:tidal forces}): 
\begin{equation}
dE_{p({\rm atm})}/dm=-G M_p/R_p.
\label{Eq : dEp atm}
\end{equation}
Numerically this converts into:
\begin{equation}
\begin{split}
& dE_{p({\rm atm})}/dm =\\
&-1.9\times 10^{13}  
\left(\frac{M_p}{M_{\rm Jup}}\right) \left(\frac{R_p}{R_{\rm Jup}}\right)^{-1} 
{\rm \ erg\,g^{-1}}.
\end{split}
\end{equation}

We note that the potential energy per unit of mass of the atmospheric gas is much higher than 
its kinetic or thermal energy, even if the gas is escaping at high velocity. 
For instance, even for gas 
escaping at high velocity of 10\,km\,s$^{-1}$, the kinetic energy is $5\times 10^{11}$\,erg\,g$^{-1}$, 
which has to be compared with the typical value of 
$-dE_{p({\rm atm})}/dm \sim 10^{13}$\,erg\,g$^{-1}$ for the known extrasolar planets
(Sect.~\ref{Diagram of the potential energy of the atmosphere and the corresponding escape rate}). 
Even for very-hot-Jupiters in which tidal forces indeed play a significant role
(Sect.~\ref{Potential energy and tidal forces}),
the potential energy is expected to be always larger than the kinetic
energy.
The kinetic energy of the escaping gas is thus neglected in the present analysis.

\subsection{Potential energy and tidal forces}
\label{Potential energy and tidal forces}

In the case of hot-Jupiters and very-hot-Jupiters, 
the gravity from the nearby parent star 
modifies the geometry of the potential energy. In those cases, the iso-potentials 
around the planet at high altitude are no more spherical and are open to the infinity when the 
potential reaches the value of the Roche lobe. Therefore the energy needed to escape the 
planet can be defined by the energy needed to reach this Roche lobe. In the case of $M_p\ll 
M_*$, the potential energy of the atmospheric gas per unit of mass is accordingly modified by 
\begin{equation}
dE'_{p({\rm atm})}/dm = dE_{p({\rm atm})}/dm + \Delta_{\rm tidal} dE_{p({\rm atm})}/dm,
\end{equation}
where (see Appendix~\ref{App:tidal forces})
\begin{equation}
\Delta_{\rm tidal} dE_{p({\rm atm})}/dm \approx
\frac{3^{4/3}}{2} \frac{ G  M_{*}^{1/3} M_{p}^{2/3}}  { a_p },
\label{Delta Ep atm}
\end{equation}
$a_p$ being the orbital distance of the planet and $M_{*}$ the mass of the central star.

The potential energy of the whole planet as defined in 
Sect.~\ref{The potential energy of the whole planet}
is also modified by tidal forces.
We can define $\Delta_{\rm tidal} E_{p}$ by
\begin{equation}
E'_{p} = E_{p} + \Delta_{\rm tidal} E_{p}.
\end{equation}
Fortunately, because Eq.~\ref{Delta Ep atm} does not depend 
on the planetary radius, the change introduced by the tidal forces 
does not depend on the planet internal structure. 
Indeed, we have 
\begin{equation}
\Delta_{\rm tidal} E_{p}=
\int_0^{R_p}\frac{3^{4/3}}{2} \frac{ G  M_{*}^{1/3} m(r)^{2/3}}  { a_p }
\frac{d  m(r)}{d  r}dr.
\end{equation}
It is easy to find that for the total potential energy of a planet we have 
\begin{equation}
\Delta_{\rm tidal} E_p \approx
\frac{3^{7/3}}{10} \frac{ G  M_{*}^{1/3} M_{p}^{5/3}}  { a_p }.
\label{Delta Ep total}
\end{equation}
Because $\Delta_{\rm tidal} dE_{p({\rm atm})}/dm$ is positive, 
the energy deposition needed to escape the planet gravity is decreased 
by the tidal forces. 
For the identified hot-Jupiters, we find that the depth of their potential well
is decreased by the tidal forces by 
typically 30 to 50\%.  
For instance, we find that 
$\Delta_{\rm tidal}  dE_{p({\rm atm})}/ |dE_{p({\rm atm})}|\sim$50\%
for OGLE-TR-56b and $\sim$33\% for HD\,209458\,b.
Keeping the hypothesis that the escape energy is mainly used to escape the potential 
well of the planet, this leads 
to a similar increase of the escape rate (Sect~\ref{Lifetime and escape rate}).

\section{Lifetime and escape rate}
\label{Lifetime and escape rate}

Assuming that the energy deposited by the EUV flux is used to extract the gas which escapes 
the potential well of the planet with an efficiency $\epsilon=1$, we can derive an 
escape rate, $dm/dt$, and a corresponding typical lifetime for a given planet. 
Assuming $-dE'_{p({\rm atm})} = \epsilon dE_{\rm EUV} $, and neglecting 
the tidal forces ($dE'_{p({\rm atm})} \approx dE_{p({\rm atm})}$), we find:
\begin{equation}
 dm/dt= -         \frac{dE_{\rm EUV}/dt}{dE_{p({\rm atm})}/dm}.
\end{equation}
From Eq.~\ref{E=E(F)}, we derive the escape rate
\begin{equation}
 dm/dt=            F_{\rm EUV}(1\,{\rm AU})
\left(\frac{\pi R_p^3}{G M_p}\right)
\left(\frac{a_p}{1\,{\rm AU}}\right)^{-2}.
\end{equation}
Numerically, this gives
\begin{equation}
\begin{split}
dm/dt= & 3.7\times 10^7
 \left(\frac{         F_{\rm EUV}(1\,{\rm AU})}{4.6\,{\rm erg\,cm^{-2}\,s^{-1}}}\right) 
\times \\
&\left(\frac{R_p}{R_{Jup}}\right)^3
\left(\frac{M_p}{M_{Jup}}\right)^{-1}
\left(\frac{a_p}{1\,{\rm AU}}\right)^{-2}
{\rm g}\,{\rm s}^{-1}.
\end{split}
\end{equation}

Concerning the lifetime, we can similarly assume 
$-E'_p=\epsilon \int_0^{t1} dE_{\rm EUV}$ to obtain a typical duration, $t_1$, 
which corresponds to the time needed to fill the whole potential well of the planet assuming 
the current energy flux. This duration can hence be called a ``lifetime'';
the nature of a
planet can change through evaporation  if $t_1$ is shorter than the typical 
lifetime of the parent star. 
If the energy received by the planet, $dE_{\rm EUV}/dt$, is assumed 
to be constant over time, from the above equation 
we derive the lifetime $t_1$:
\begin{equation}
 t_1=\frac{-E'_p}{         dE_{\rm EUV}/dt}.
\end{equation}

However, when the planetary mass evolves, the radius and the corresponding 
energy flux captured by the planet changes, consequently also the real lifetime changes. 
A more detailed analysis is possible by considering the variations of the escape rate 
as a function of time. 
This leads to a new ``lifetime'', $t_2$, which can be defined by 
\begin{equation}
\begin{split}
& -E'_p =  \\
& \left(\frac{a_p}{1\,{\rm AU}}\right)^{-2} 
\int_{t_0}^{t_2}  \pi R_p(t')^2 
 (         F_{\rm EUV}(1\,{\rm AU})(t'))
dt',
\end{split}
\end{equation}
where $t_0$ is the initial time of the system.
At this stage a useful quantity is the mean value of the EUV energy flux from the central star 
$<dE_{\rm EUV}/dt>$ defined by:
\begin{equation}
\begin{split}
& <dE_{\rm EUV}/dt>(t) \equiv \\
& \left(\frac{a_p}{1\,{\rm AU}}\right)^{-2}
\times 
\frac{\int_{t_0}^{t} \pi R_p(t')^2 (F_{\rm EUV}(1\,{\rm AU})(t')) dt'}
{(t-t_0)},
\end{split}
\label{<dE/dt>}
\end{equation}
which includes the variations of the EUV flux 
and the variation of the planetary radius which both are larger during the 
early stage of the
planetary system (Guinan \& Ribas 2002; Ribas et al.\ 2005; Burrows et al.\ 2000).
We can define a correction factor for time variation of the energy flux
received by a planet, 
\begin{equation}
\gamma\equiv \frac{<dE_{\rm EUV}/dt>}{dE_{\rm EUV}/dt}.
\label{gamma}
\end{equation}
Hence we have
\begin{equation}
t_2 \approx \gamma^{-1} t_1 .
\end{equation}

We now have to evaluate the impact of this $\gamma$ correction factor. First,
the variation of the EUV flux as a function of time has been thouroughly analyzed for G stars 
by Ribas et al.\ (2005) within the ``Sun in time'' program. A recent work reports similar results for K type
stars (Lakatos et al.\ 2005). Roughly speaking, the EUV flux can be assumed to follow a power law
in the form 
$F\propto t^{-\alpha}$, where $\alpha$ is close to~1. For G type stars, the 1-1200\AA\ EUV flux is
found to have $\alpha\sim 1.23$, while the Lyman-$\alpha$ has $\alpha=0.72$ (Ribas et al.\ 2005). 
For K type 
stars, the stellar C\,{\sc iv} emission lines, which are believed to emerge from the same coronal
region as the EUV flux, follows a power law with $\alpha\sim 0.94$. We therefore can consider 
a power law with $\alpha=1$ as a good approximation (see also Vardavas 2005). 
Assuming that the increase of the UV luminosities 
in the past has to be limited and should ``saturate'', we consider that the luminosities do not
vary at ages below 100 million years (Lakatos et al.\ 2005). With these values, and neglecting for the moment 
the variation of the planetary radius, 
we find that for $\alpha= 1.23$ we have $\gamma=6.5$ at $t=5$\,Gyr, 
and for $\alpha=0.75$ we have $\gamma =4$. 
We can therefore consider a power law with $\alpha=1$ as a good approximation.

The variation of the planetary radius is a more complex matter because this depends 
on the planetary characteristics like mass and orbital distance. Nonetheless, 
empirically
we found that the published time variation of the planetary radius 
(Burrows et al.\ 2000, Guillot 2005) 
can be well fitted by a simple scaling law
in the form :
\begin{equation}
R_p(t) = R_{\infty} + \beta t^{-0.3}.
\label{R_p(t)}
\end{equation}

We find good fits  with $\beta =0.2 R_{\infty}\,{\rm Gyr}^{0.3}$ for $M_p \ga 0.3\,M_{\rm Jup}$,
and with $\beta =0.3 R_{\infty}\,{\rm Gyr}^{0.3}$ for $M_p \approx 0.1\,M_{\rm Jup}$.

Using $\alpha=1$ for the power low of the luminosity variations and $\beta =0.2 R_{\infty}\,{\rm Gyr}^{0.3}$
for the variation of the planetary radius, we find $\gamma = 6.8$ at $t=2$\,Gyr 
(the typical age of the F6-F7 stars harboring planets, Saffe et al.\ 2005) 
and $\gamma = 8.0$ at $t=5$\,Gyr (taken as the typical age of later type
stars). 
Using for less massive planets $\beta =0.3 R_{\infty}\,{\rm Gyr}^{0.3}$, we find 
$\gamma = 8.7$ at $t=2$\,Gyr  and $\gamma = 10.0$ at $t=5$\,Gyr. 
In the following we consequently decided to consider $\gamma=7$ for F6-F7 stars, and $\gamma=8$ 
for later type stars. 

If we consider that the EUV luminosities can be extrapolated with the $t^{-1}$ function 
in the past earlier than at the age of 100~million years, down to the age of 30 million years, 
$\gamma$ is then increased by about 25\%. We thus conclude that 
the correction factor for time variation of the energy flux does not strongly depend 
on the epoch of this ``saturation'' of the EUV luminosities in the past.

\section{The energy diagram}
\label{The energy diagram}

\begin{figure}[bth!]
\begin{center}
\psfig{file=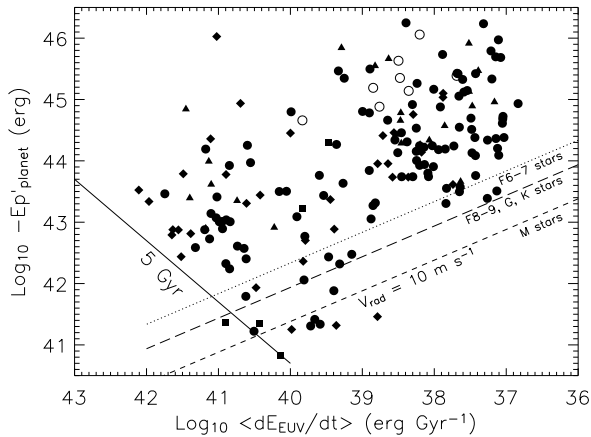,width=\columnwidth,angle=90} 
\caption[]{
Plot of the potential energy of the extrasolar planets as a function of the 
mean EUV energy received per billion of years, $<dE_{\rm EUV}/dt>$.
To keep the planets with the smallest orbital distances 
in the left part of the diagram, the direction of the abscissa axis is chosen 
with the largest value of the mean energy flux toward the left.
The 182 identified planets are plotted with symbols depending on the type of the
central star: triangles for F stars, filled circles for G stars, diamonds for K stars
and squares for M stars; planets orbiting class III stars are plotted with empty circles.     
The absence of planets in the bottom right region can be explained by the detection 
limit of the radial velocity surveys. 
There are indeed very few planets with radial velocities 
of the parent stars below the lines for which the radial velocity is
10\,m\,s$^{-1}$. 
The radial velocity iso-curves at 10\,m\,s$^{-1}$ 
have been plotted for various stellar types (see Appendix~\ref{App:rad vel}); 
the iso-curves for F8-F9, G and K-type stars are found to be nearly superimposed 
and are plotted with a single line. 

The striking result is the absence of planets in the bottom left region 
which corresponds to light planets (small $-E'_p$) at short orbital distances 
(large $<dE_{\rm EUV}/dt>$).
A plot of the lifetime line at $t_2$=5\,Gyr, shows that there are no planets in
this part of the diagram simply because this is an evaporation-forbidden region.
Planets in this region would receive more EUV energy than needed to fill the
potential well of the planet, and evaporate in less than 5\,Gyr.
\label{Ep_planet_vs_Euv}}
\end{center}
\vspace{-0.7cm}
\end{figure}

With the equations derived in the previous sections, 
we can now plot diagrams of the potential energy versus energy flux 
(Fig.~\ref{Ep_planet_vs_Euv}).
In particular Eqs.~\ref{E_p} and~\ref{Delta Ep total} are used 
to derived the potential energy, $E'_p$, 
and Eq.~\ref{<dE/dt>} (or a combination of Eqs.~\ref{E=E(F)} and~\ref{gamma})
is used to derive the mean EUV energy 
received per unit of time by the planets, $<dE_{\rm EUV}/dt>$.
We used the characteristics of the 182 identified planets as tabulated
by Schneider (2006) on June 15$^{\rm th}$, 2006.

The only remaining difficulty in the drawing of the plot lies in the mass determination.
Except for the case of transiting planets (Sect.~\ref{Uncertainties}), 
most of the planets have only a determination of $M_p\sin i$, 
where $i$ is the orbit inclination. We compensate for this effect by
multiplying the $M_p\sin i$ by a factor $\sqrt{2}$. This $\sqrt{2}$ factor 
is taken because the real mass $M_p$ is statistically smaller than 
$\sqrt{2} M_p\sin i$ in 50\% of the cases, and larger in 50\% of the cases. 
The result is thus a spread of the position of the planets in the energy diagram.
For transiting planets because $\sin i\approx 1$, we keep the measured mass without
applying a correction factor.

From these planetary masses and using the tabulated types of the parent stars 
we estimate the planetary radius
by fitting the radius versus mass curves calculated by Guillot (2005).
This model assumes a central rock core and a gaseous 
envelope following an hydrogen-helium equation of state with an helium 
mixing ratio $Y=0.30$. The estimated radii should not 
be sensitive to the evaporation, because they result from a calculation
of the planet interior at several bars pressure not affected by the evaporation 
operating much higher in the atmosphere; this is
confirmed by the absence of correlation of the observed radii
of transiting planets with age and EUV flux received
during the planets' lifetime (Guillot et al.\ 2006).
An indetermination remains for the very low mass planets (Neptune mass planets)
because their radius (or density) as a function of their mass 
strongly depends on their nature. As a first approach and to plot the energy diagram
in Fig.~\ref{Ep_planet_vs_Euv}, we consider that these planets have 
a typical density of 2\,g\,cm$^{-3}$ for 15 Earth mass and 
6\,g\,cm$^{-3}$ for 6 Earth mass. A detailed analysis of these planets 
in this diagram will be presented in Sect.~\ref{Neptune mass planets in the diagram}.

The first plot (Fig.\ref{Ep_planet_vs_Euv}) 
shows the total potential energy of the
planet as a function of the mean energy flux $<dE_{\rm EUV}/dt>$.
We decided to plot the mean energy flux on the abscissa axis in reverse order  
with larger values in the left. This choice is done to keep the planets with
short orbital distances in the left part of the diagram for consistency with previous 
plots like, for instance, the diagram of Lecavelier des Etangs et al.\ (2004).
The opposite of the potential energy, $-E'_p$, is larger for more massive planets,
and massive planets are therefore found in the upper part of the diagram.
In some way this diagram is close to the classical diagram of mass 
versus orbital distance. 
However, in the present diagram the position reflects the
impact of the evaporation on the planetary nature. 

We plot the diagram with the 182 identified extrasolar planets.
The striking result is the absence of planets in two different regions of the diagram. 
The absence of planets in the bottom right region can be explained by the detection 
limit of the radial velocity surveys. We plotted the limit for radial velocity of
10\,m\,s$^{-1}$. As expected, we see that very few planets have radial velocities 
of the parent stars below this limit.

The most important region devoid of planets in the diagram is the bottom left region.
This region should contain light planets at short orbital distances. However
a plot of the lifetime line at $t_2$=5\,Gyr, shows that there are no planets in
this part of the diagram simply because this is an evaporation-forbidden region.
Planets in this region would receive more EUV energy than needed to fill the
potential well of the planet and evaporate the gas, leaving a remaining core,
an evaporation remnant (also named a ``chthonian'' planet; Lecavelier des Etangs 
et al.\ 2004).

From the position in the diagram, the typical lifetime of a given planet can be 
rapidly extracted. If the mean energy flux $<$$dE_{\rm EUV}/dt$$>$
is given in unit of {\it erg per billion years}, 
and the potential energy is given in unit of {\it erg}, 
the simple ratio of both quantities provides the
corresponding lifetime in billion of years. 
In the diagram, lifetime isochrones are straight lines.
In the Fig.~\ref{Ep_planet_vs_Euv}, we plotted
the line corresponding to the lifetime of 5\,Gyr. 

\section{Diagram of the potential energy of the atmosphere and the corresponding escape rate}
\label{Diagram of the potential energy of the atmosphere and the corresponding escape rate}

\begin{figure}[tb!]
\begin{center}
\psfig{file=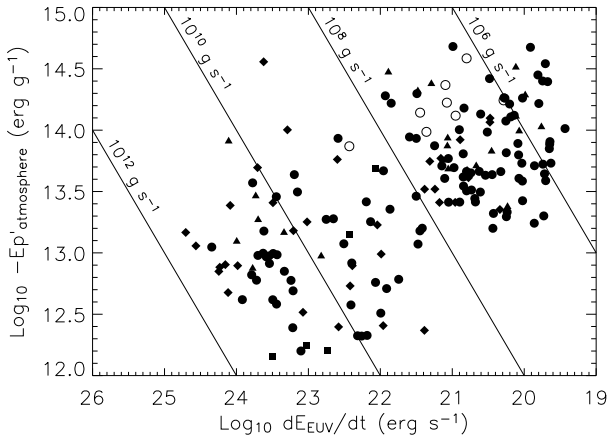,width=\columnwidth,angle=90} 
\caption{Plot of the potential energy per unit of mass of the atmosphere 
of the extrasolar planets as a function of the EUV input flux. 
Unlike Fig.~\ref{Ep_planet_vs_Euv}, the potential energy is 
per unit of mass at one planetary radius in erg\,g$^{-1}$; and the EUV input flux is
the value at the present time in erg\,s$^{-1}$. This allows to directly estimate
the corresponding escape rate at the present time in g\,s$^{-1}$.  
The escape rates are drawn with straight lines for $10^6$, 
$10^8$, $10^{10}$ and $10^{12}$\,g\,s$^{-1}$. 
The 182 identified planets are plotted with the
same symbols as in Fig.~\ref{Ep_planet_vs_Euv}.
\label{Ep_atm_vs_Euv}}
\end{center}
\vspace{-0.7cm}
\end{figure}

With the above equations 
(Eqs.~\ref{E=E(F)}, \ref{Eq : dEp atm} and~\ref{Delta Ep atm})
we can also plot the
potential energy per unit mass of the atmosphere as a function of the 
present time energy flux (Fig.~\ref{Ep_atm_vs_Euv}).
We again superimposed the position of the 182 identified planets.

Although apparently similar to the previous diagram, this new diagram allows the
estimate of the actual evaporation rate for a given planet, 
assuming that most
of the energy in the extreme ultraviolet is used to escape the combined 
effect of planet gravity and tidal forces.
In Fig.~\ref{Ep_planet_vs_Euv}, we considered the total potential energy 
providing a hint on the total escape mass and time. Here in 
Fig.~\ref{Ep_atm_vs_Euv}, we consider the potential energy \emph{per unit of mass}
(erg\,g$^{-1}$) of the atmospheric gas versus the \emph{present} time input energy
(erg\,s$^{-1}$). The ratio of both quantities provides a hint of the 
present escape \emph{rate} (g\,s$^{-1}$).

We see in the plot that the extrasolar planets must have escape rate in 
the range $10^{5}$ to $10^{12}$\,g\,s$^{-1}$. The largest escape rates are obtained
for OGLE-TR-113\,b, HD\,189733\,b, HD\,46375\,b, HD\,63454\,b, and TrES-1,
for which we obtain escape rates of 
$3.5\times 10^{11}$, $3.2\times 10^{11}$, $2.8\times 10^{11}$, $2.5\times 10^{11}$, 
and $2.2\times 10^{11}$\,g\,s$^{-1}$, respectively.
These five stars are cataloged as K type stars and therefore assumed to
have large extreme ultraviolet luminosities. The next largest escape rate 
is found for an M type star, GJ \,436\,b, for which we estimate a maximum
escape rate of the order of 
$2.2\times 10^{11}$\,g\,s$^{-1}$.

For HD\,209458\,b, 
we obtain an escape rate of $9\times 10^{10}$\,g\,s$^{-1}$, which is 
consistent with the lower limit of $\sim10^{10}$\,g\,s$^{-1}$ obtained
from the interpretation
of the Lyman-$\alpha$ transit light curve 
(Vidal-Madjar et al.\ 2003; Vidal-Madjar \& Lecavelier des Etangs 2004).

\section{Uncertainties}
\label{Uncertainties}

For the observed planets, the position in the previous diagrams are obtained by simple estimates of the
EUV luminosities and planetary radii. These rough estimates are subject to uncertainties. These uncertainties
add some dispersion in the clouds of points of a given diagram; however with a limited dispersion, 
this does not severely affect 
the main trends and the corresponding statistical conclusions, like the empty region observed for 
lifetime below 5\,Gyr. More critically, when we consider a particular planet, we need to have 
an estimate of the uncertainties introduced by above approximations.

\begin{figure}[bt!]
\begin{center}
\psfig{file=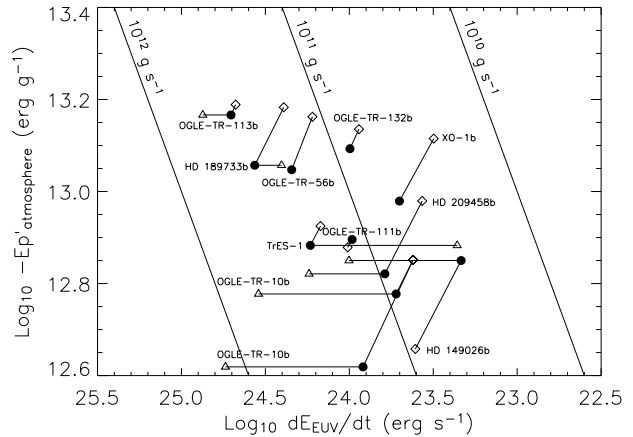,width=\columnwidth,angle=90} 
\caption{Plot of the potential energy per unit of mass of the atmosphere 
of the ten
\emph{transiting} planets as a function of the EUV input flux. 
For each planet the position is calculated using the
radius as modeled (diamonds) and the radius as effectively measured (circles). 
For OGLE-TR-10b we used the two possible planetary radii quoted by Santos et al.\ 2006. 
When available, the EUV luminosities estimated from the stellar 
rotational velocity are plotted with triangles.
\label{Ep_atm_vs_Euv_transit}}
\end{center}
\vspace{-0.7cm}
\end{figure}

To evaluate the impact of these approximations on the planetary radius and stellar EUV luminosities, 
we used the
transiting planets. For these planets, we have a measured radius and also a better estimate of 
the EUV luminosities through 
the rotational velocity of the central stars $v\sin i$ 
(when measured, in few cases only upper limits are available). 
Because the rotational axis of the stars is expected to be close to the 
orbital axis of the planets (see for instance Queloz et al.\ 2000), 
and for transiting planets $\sin i \cong 1$, the measured
$v\sin i$ are in fact true rotational velocities 
from which better estimates of the EUV luminosities can be obtained.
We plotted these transiting planets in the 
diagram of the potential energy per unit of mass versus 
EUV energy flux (Fig.~\ref{Ep_atm_vs_Euv_transit}). 

Concerning the radius estimate, OGLE-TR-10b is a special case because this planets has presently
two different radius estimates which are significantly different. 
From the analysis of Bouchy et al.\ 2005a, Santos et al.\ 2006 obtain a radius of 
1.43 Jupiter radii, while they obtain a radius of 1.14 Jupiter radii using 
the new observational data of Holman et al.\ 2005. Because we cannot discriminate
between both results, we decided to keep both radii in the present discussion. 

We first evaluate the uncertainty on the radius estimate and its impact on the evaporation rate. 
Because $dE_{p({\rm atm})}/dm \propto R_p^{-1}$ and $dE_{\rm EUV}/dt \propto R_p^2$, the resulting 
escape rate $dm/dt$ is quite sensitive to the planetary radius and is proportional
to $R_p^3$. For each planet in Fig.~\ref{Ep_atm_vs_Euv_transit}, we consider the
radius obtained from our fit to Guillot's models (Guillot 2005; plotted with diamonds) 
and the radius as effectively measured (plotted with filled circles). 
In the ten 
transiting planets the difference between the two corresponding escape rates
has a mean value of 0.13~dex and a dispersion of 0.22~dex. The largest differences
are obtained for OGLE-TR-10 (increase of the escape rate by 0.45~dex with the radius 
given by Bouchy et al.\ 2005a), 
HD\,209458\,b (increase of the escape rate by 0.33~dex with the real radius)
and HD\,149026\,b (decrease of the escape rate by 0.41 dex with the real radius).
This is due to the unexplained excess of planetary radius of OGLE-TR-10 and
HD\,209458\,b (Guillot et al.\ 2005, Laughlin et al.\ 2005) 
and the small radius of HD\,189733\,b explained by its
large metallicity and its suspected massive core (Sato et al.\ 2005).
To conclude, although the escape rate is quite sensitive to the radius estimate
calculated from the planet characteristics, the transiting planets allows us to
estimate that this leads to a typical uncertainty of less than 0.3~dex in the calculated
escape rate.

The impact of the EUV energy flux estimate is more difficult to constrain
as we do not have real EUV flux measured for any of the stars harboring
planets. However, we can use another estimate of the EUV luminosities through 
true rotational velocity when available in the published literature 
({\it i.e.}\ for six of the ten transiting planets).
For these six planets we calculated the EUV luminosities 
from their rotational velocity using the results of Wood et al.\ (1994).
Using {\sl Rosat} S2 measurements of stars within 10~parsecs 
Wood et al.\ (1994) found a close correlation between the S2 luminosity, $L_{S2}$, and 
the stellar rotational velocity, $v_{\rm rot}$: $L_{S2}\propto v_{\rm rot}^{1.4}$. 
From this equation, and assuming 
$F_{\rm EUV}=4.6$\,erg\,cm$^{-2}$\,s$^{-1}$ at 1\,AU and $v_{\rm rot}=2.0$\,km\,s$^{-1}$ 
for the Sun, 
we find an EUV energy flux of:
\begin{equation}
F_{\rm EUV}(1\,{\rm AU}) =  4.6 
\left(
 \frac{v_{\rm rot}}{2.0\,{\rm km\,s}^{-1}}
                       \right)^{1.4} 
{\rm \,erg\,cm^{-2}\,s^{-1}}
\label{Euv(vrot)}
\end{equation}
at a distance of 1\,AU from a star with a rotational velocity $v_{\rm rot}$.

For the six transiting planets with published rotational velocity we plot
the result in Fig.~\ref{Ep_atm_vs_Euv_transit} with triangles.
In these six planets the difference between the two corresponding escape rates
has a mean value of 0.18~dex and a dispersion of 0.60~dex. The biggest differences
are obtained for OGLE-TR-10 (increase of the escape rate by 0.82~dex with the EUV luminosity
estimated from the rotational velocity) and TrES-1 
(decrease of the escape rate by 0.87 dex).
This is due to the large rotational velocity of OGLE-TR-10 with $v \sin i$= 7.7\,km\,s$^{-1}$
(Bouchy et al.\ 2005a), 
and the small rotational velocity of TReS-1 with $v \sin i$= 1.08\,km\,s$^{-1}$
(Laughlin et al.\ 2005).
As a result, we see that in two of the six cases, we have a large uncertainty of the EUV luminosity
with consequently a relatively large uncertainty in the escape rate ($\sim 0.8$~dex),
for the four other cases the dispersion is reduced to 0.4~dex.
In general, taken into account the two uncertainties (radius and energy flux
estimates), we find that the differences in the estimated escape rates
have a typical dispersion of 0.7~dex. 

\section{Evolution tracks}
\label{Evolution tracks}

The diagram described above aims at describing the possibility of the planets evolution on billion years
time scale through the evaporation process. The position in the diagram can be calculated
or evaluated at a given time, in particular in the present epoch (assuming an age for the parent star).
It is thus interesting to see how a planet can evolve in such a diagram. An interesting
aspect of the diagram is that the position in the diagram in itself 
contains the information of the evolution timescale.

\begin{figure}[tbh!]
\begin{center}
\psfig{file=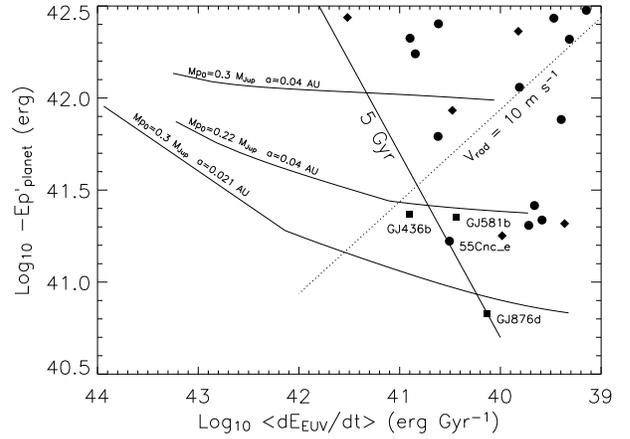,width=\columnwidth,angle=90} 
\caption{Plot of the potential energy of the extrasolar planets 
as a function of the EUV input flux with evolutionary tracks superimposed.
This is an enlargement of Fig~\ref{Ep_planet_vs_Euv}.
Time evolution of planets at constant orbital distance $a_p=0.021$\,AU and
$a_p=0.04$\,AU, corresponding to the orbital distances of GJ\,876\,d and two Neptune-mass
planets (GJ\,581\,b and 55\,Cnc\,e), respectively. GJ\,581\,b and 55\,Cnc\,e
are on the track of a planet with an initial mass $M_p(t_0)$=0.22\,$M_{\rm jup}$.
GJ\,876\,d is on the track of a planet with an initial mass of  
$M_p(t_0)\sim $0.3\,$M_{\rm jup}$, 
showing that it can be the remnant of a former hot Saturn-mass planet.
\label{Ep_planet_vs_Euv_track}}
\end{center}
\vspace{-0.7cm}
\end{figure}

We thus calculated evolution tracks for few cases of planets for which the initial conditions
imply rapid evolution through evaporation. This concerns light planets at
short distances (Saturn mass planets at very-hot-Jupiter distances).
In calculating these evolution tracks, the critical parameter is found to be
the planetary radius and mass at a given age, both evolving with time. 
Again we assumed that the EUV flux varies as $t^{-1}$ and 
used our fit to the Guillot et al.\ models 
and the corresponding time evolution described by Eq.~\ref{R_p(t)}. 
The result is plotted in Fig.~\ref{Ep_planet_vs_Euv_track}.

We calculated the time evolution of planets at orbital distance $a_p=0.021$\,AU and
$a_p=0.04$\,AU, corresponding to the orbital distances of GJ\,876\,d and two Neptune-mass
planets (GJ\,581\,b and 55\,Cnc\,e), respectively. We find that GJ\,581\,b and 55\,Cnc\,e
can well be the remnant of a planet with an initial mass of 0.22\,$M_{\rm jup}$. The evolution
track of a planet with an orbital distance of 0.021\,AU and an initial mass of 0.3\,$M_{\rm jup}$ 
shows that GJ\,876\,d can be the remnant of a planet whose initial mass could be up to the mass
of Saturn.

\section{Neptune mass planets in the diagram}
\label{Neptune mass planets in the diagram}

\begin{figure}[tb!]
\begin{center}
\psfig{file=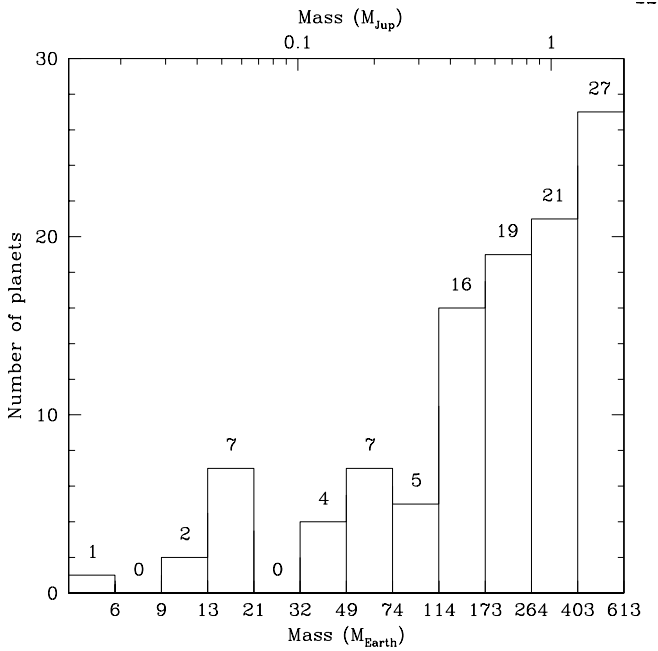,width=\columnwidth} 
\caption{Mass distribution of planets discovered through radial velocity. 
\label{Mass_Histo}}
\end{center}
\vspace{-0.7cm}
\end{figure}

A plot of the mass distribution of the extrasolar planets shows that Neptune mass planets play
a particular role (Fig.~\ref{Mass_Histo}). 
Ten planets have been found with mass below 0.067\,$M_{\rm jup}$ 
(21~Earth-mass), 
while no planets have been identified with mass between 0.067\,$M_{\rm jup}$ and 
0.11\,$M_{\rm jup}$ (35~Earth-mass). 
The presence of this gap is not related to a bias in the radial velocity searches, 
since more massive planets are easier to detect (at the same orbital distances). 
This reveals the different nature of these
Neptune mass planets orbiting at short orbital distances. But their nature is still a matter of 
debate (Baraffe et al.\ 2005). In particular the question arises 
if they can be the remnants of evaporated more massive
planets (``chthonian planets'') as foreseen in Lecavelier des Etangs et al.\ (2004) ?
Other possibilities include gaseous Neptune-like planets, super-Earth (Santos et al.\ 2004) 
or ocean-planets (Kutchner 2003, L\'eger et al.\ 2004).

\begin{figure}[tbh]
\begin{center}
\psfig{file=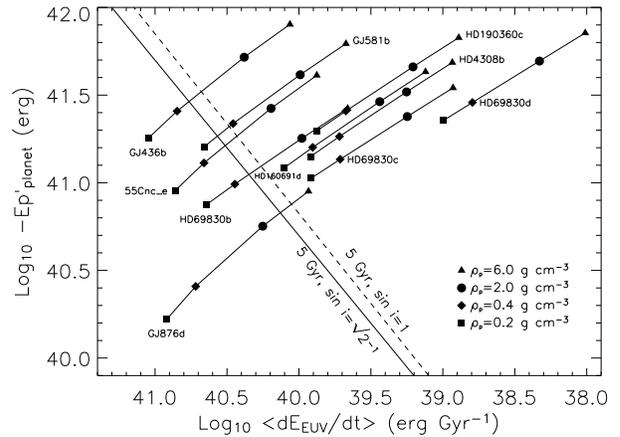,width=\columnwidth,angle=90} 
\caption{Plot of the potential energy of the Neptune mass planet 
as a function of the EUV flux for various planets' density. 
For GJ\,581\,b, GJ\,436\,b, HD\,69830\,b, 55\,Cnc\,e and GJ\,876\,d, and assuming 
$\sin i=\sqrt{2^{-1}}$, lifetime shorter than 5\,Gyr are obtained 
for densities below 0.28, 0.55, 0.56, 0.69 and 3.1\,g\,cm$^{-3}$, respectively.
If $\sin i$=1 (dotted line), the critical (minimum) densities are 
increased to 0.38, 0.74, 0.78, 0.93 and 4.2\,g\,cm$^{-3}$, 
for GJ\,581\,b, GJ\,436\,b, HD\,69830\,b, 55\,Cnc\,e, and GJ\,876\, respectively.
 \label{Ep_planet_vs_Euv_HN}}
\end{center}
\vspace{-0.7cm}
\end{figure}

The mass of these objects being measured, the freedom for their position in the
energy diagram mainly relies on the determination of their radius, 
or, in other words, of their mean density, 
\mbox{\it i.e.}, their nature.
As a first approach, in Fig.~\ref{Ep_planet_vs_Euv} we considered 
that these planets have a typical density of 2\,g\,cm$^{-3}$ for 15~Earth mass 
and 6\,g\,cm$^{-3}$ for 6~Earth mass. 
Now we can test the density hypothesis and try to conclude on the possible
nature of these planets.
We plotted the position of these Neptune mass planets in the energy diagram 
with different hypothesis on their density (Fig.~\ref{Ep_planet_vs_Euv_HN}). 
We used mean planetary density of $\rho_p$=6\,g\,cm$^{-3}$ for a typical density of 
refractory-rich planets which should describe the chthonian and super-Earth planets. A lower density 
on the order of $\rho_p$=2\,g\,cm$^{-3}$ can be considered as more plausible for volatile-rich
planets describing the ocean planets. For gas-rich planets we assumed much lower density
of $\rho_p$=0.2\,g\,cm$^{-3}$ and $\rho_p$=0.4\,g\,cm$^{-3}$, describing planets which should
look more like irradiated Neptune-like planets 
(Baraffe et al.\ 2006 find densities of 0.1 to 0.25\,g\,cm$^{-3}$ for the final stage of 
an evaporated gaseous planets with a 6 Earth mass core and final mass of about one Neptune mass).

In the energy diagram, HD\,190360\,c, HD\,4308\,b, HD\,160691\,d, HD\,69830\,c and HD\,69830\,d 
are all located in
the possible domain with lifetime above 5\,Gyr, independently of their assumed density. 
Even with very low densities down to 0.2\,g\,cm$^{-3}$, these planets do not receive enough 
energy from their central star to evaporate significantly.
For GJ\,581\,b only the extreme case $\rho_p$=0.2\,g\,cm$^{-3}$ 
is in the evaporation-forbidden region.
Owing to their orbital distances between 0.09 and 0.63\,AU,
it appears that the evaporation is not efficient enough 
to modify the nature if these planets. 
These planet are not likely to be the evaporation remnants of previously more massive planets 

For GJ\,436\,b, 55\,Cnc\,e and HD\,69830\,b, density below 0.4\,g\,cm$^{-3}$ would
place these planets in the evaporation-forbidden region of the energy diagram. 
If these planets were formerly low density (gaseous giant) planets, they must have 
lost a large fraction of their upper atmospheric gas. We therefore conclude that, 
with low orbital distances in the range 0.03 to 0.08\,AU, 
these planets cannot belong to the category of low mass gaseous planet
(irradiated Neptune-like planets). 
The position of these planets in the energy diagram shows that they must contain 
a large fraction of solid/liquid material in the form of volatile or refractory elements. 
These planets can be the remnants of previously more massive planets (chthonian planets) 
or alternatively volatile-rich planets (ocean planets) or rocky planets (super-Earths).

It is noteworthy that the conclusion for HD\,69830\,b is consistent with
the one reached by Lovis et al.\ (2006) who concluded from
formation and migration models that it must have a
mainly rocky composition.

For GJ\,876\,d, even the density of an ocean planet 
creates some problem. 
GJ\,876\,d needs to be a planet dense enough to be located
above the $t_2=5$\,Gyr lifetime limit. This planet requires
a density larger than 3.1\,g\,cm$^{-3}$ for its atmosphere to survive. 
GJ\,876\,d could be a big rocky planet, like a super-Earth, or a refractory remnant of
a previous more massive planet (an evaporated ocean planet?).
However, we note that its parent star is an M star, and we have very little information
about the EUV flux of non active M dwarfs; even if every M star has period of strong
activity remains unclear. A more detailed analysis of the energy flux from this
star and its possible consequences on this intriguing planet is needed.

In brief, the energy diagram allows us to trace three different categories for
the ten presently identified Neptune mass planets. For six of them, 
the EUV input energy seems not strong enough to affect significantly these 
planets; we cannot conclude on their nature. 
For three other planets, it appears that
they cannot be a kind of low mass gaseous planets. These planets must contain 
a large fraction of solid/liquid material. Finally, GJ\,876\,d must be 
dense enough to survive the strong EUV energy flux from its nearby parent 
star. This planet must contain a large fraction of massive elements.

\section{Conclusion}

To describe the evaporation status of the extrasolar planets, we proposed to consider
an energy diagram in which the potential energy of the planets is plotted versus the
energy received by the upper atmosphere. We presented a basic method to estimate
these quantities. For the energy flux, we proposed a first approach
using the type of the parent stars or, if available, their
rotation period. For the potential energy, we included the modification 
of the gravity field by the tidal forces from the parent stars. 
This description allows a quick estimate of both the escape rate of the 
atmospheric gas and the lifetime of a planet against the evaporation process. 
We found that the region of lifetime below 5\,Gyr is devoid of planets, 
which reinforces the belief that this diagram
is an appropriate description of the evaporation.
In the evaporation-forbidden region, the evaporation can 
modify the nature and the characteristics of the planets 
such that they move to region of the diagram corresponding to longer lifetime 
and consequently become long-lived.

The escape rates estimated from the diagram are found to be in the range
$10^{5}$\,g\,s$^{-1}$ to $10^{12}$\,g\,s$^{-1}$, 
with few cases above $10^{11}$\,g\,s$^{-1}$. 
Although this is a crude estimate, the estimated 
escape rate for HD\,209458\,b is found to be consistent 
with the lower limit of $10^{10}$\,g\,s$^{-1}$ obtained from interpretation of
the H\,{\sc i} Lyman-$\alpha$ observations.

Finally, this diagram allows the description of the possible nature of the recently
discovered Neptune-mass planets. We found that GJ\,436\,b, 55\,Cnc\,e and 
HD\,69830\,b cannot be low mass gaseous planets. These planets must contain 
a large fraction of solid/liquid material. Concerning GJ\,876\,d, we argued that
it must be dense enough to survive the strong EUV energy flux from its nearby parent 
star. GJ\,876\,d must contain a large fraction of massive elements.

\begin{acknowledgements}
 This work has been motivated by many discussions with various people. 
 In particular, I am extremely grateful to F.~Bouchy, G.~H\'ebrard, J.~McConnell, F.~Pont,
 G.~Tinetti and A.~Vidal-Madjar.  
 I warmly thank the anonymous referee for his detailed comments which helped 
 to clarify the paper. 
 I thank the International Space Science Institute (ISSI) in Berne, Swizerland, 
 for support and discussion during the workshop on transiting extrasolar planets.
\end{acknowledgements}

\appendix
\section{The potential energy of a gaseous planet}
\label{App:Ep}

The potential energy of a gaseous sphere of radius $R_p$ is given by 
\begin{equation}
E_p=-G\int_0^{R_p}\frac{m(r)}{r}\frac{d  m(r)}{d  r}dr
\label{App Eq Ep int}
\end{equation}
where $G$ is the gravitational constant and $m(r)$ is the mass inside a radius $r$.
The internal structure, characterized by $m(r)$, is given by 
\begin{equation}
\frac{d  P }{d  r}= - \rho(r) \frac{Gm(r)}{r^2},
\label{Eq dP/dr}
\end{equation}
where $\rho(r)$ is the density given by 
\begin{equation}
\frac{d  m }{d  r}=4\pi r^2 \rho(r),
\label{dm/dr}
\end{equation}
and the relation between the pressure $P(r)$ and the density $\rho(r)$
is given by the equation of state.
An analytical solution can be found if we assume that 
the equation of state follows a polytropic with an 
index $n=1$, which is found to be a good approximation for 
the gaseous planets interior (de Pater \& Lissauer 2001).
In that case, the equation of state is 
\begin{equation}
P= K \rho^2,
\label{Eq P(rho)}
\end{equation}
where $K$ is a constant. The equations~\ref{Eq dP/dr}, \ref{dm/dr} and \ref{Eq P(rho)} 
can be analytically solved. This gives
the density profile within the planet (de Pater \& Lissauer 2001):
\begin{equation}
\rho(r)=\rho_c \left(\frac{\sin(C_K r)}{C_K r}\right),
\label{Eq rho(r)}
\end{equation}
with 
\begin{equation}
C_K=\frac{\pi}{R_p}=\sqrt{\frac{2\pi G}{K}}.
\label{CK}
\end{equation}
Putting Eq.~\ref{Eq rho(r)} into Eq.~\ref{App Eq Ep int}, this gives
\begin{equation}
E_p=-\frac{12G\pi^3\rho_c^2}{C_K^5}.
\label{first Ep}
\end{equation}
Also using Eqs.~\ref{dm/dr}, \ref{Eq rho(r)} and~\ref{CK}, the planet mass, $M_p$ is given by 
\begin{equation}
M_p=m(R_p)=m(\pi/C_K)=\frac{4\pi^2\rho_c}{C_K^3}.
\label{Eq Mp}
\end{equation}
Finally, putting Eq.~\ref{CK} and \ref{Eq Mp} into~\ref{first Ep} we find
the potential energy of a gaseous sphere in which the equation of state 
follows a polytropic with an index $n=1$:
\begin{equation}
E_p=-\frac{3}{4}G M_p^2/R_p.
\end{equation}

\section{Modification of the potential energy by tidal forces}
\label{App:tidal forces}

In the field of a planet of mass $M_p$, at a distance $r$ from
the planet center, the potential energy field, $\chi$, follows the equation
\begin{equation}
\chi(r)=-\frac{G M_p}{r} + C,
\label{Eq chi}
\end{equation}
where $G$ is the gravitational constant and $C$ is a constant. 
The potential energy of a unit mass of the atmospheric gas of a planet, 
$dE_{p({\rm atm})}/dm$, can be defined 
by the opposite of the energy needed to
escape the planet gravity, in other words to reach infinite distance from
the planet:
\begin{equation}
dE_{p({\rm atm})}/dm\equiv \chi(R_p)-\chi(\infty)=-G M_p/R_p.
\label{App Eq Ep atm}
\end{equation}
where $R_p$ is the radius of the planet.
This demonstrates the well known Eq.~\ref{Eq : dEp atm} given 
in Sect.~\ref{The potential energy per unit of mass in the atmosphere}.

However, in a star-planet system, the potential energy field is modified
by the star gravity and centrifugal forces. 
Along the star-planet axis, the potential energy field, $\chi'$, follows the equation:
\begin{equation}
\begin{split}
\chi'(r)= & \chi(r)+ \Delta \chi (r) \\
= & - G (M_p+M_*) \left[ \frac{\mu}{r} 
                     + \frac{1-\mu}{a_p-r} 
                     + \frac{\left[\left(1-\mu\right) a_p-r\right]^2}{2 a_p^3} 
                \right]\\
  & + C,
\label{Eq chi'}
\end{split}
\end{equation}
where $r$ is the distance from the planet center in the direction of the
star on the star-planet axis, $a_p$ is the star-planet distance, 
$M_*$ is the stellar mass and $\mu$ is defined by
\begin{equation}
\mu=\frac{M_p}{M_p+M_*}.
\end{equation}
In the case $r \ll  a_p$ and $\mu \ll 1$, we can derive
\begin{equation}
\begin{split}
\frac{\Delta\chi(r)}{G (M_p+M_*)}= 
              - \frac{3}{2a_p}\left(
                        1+\frac{r^2}{a_p^2}
                        +O(\frac{r^3}{a_p^3}) + O(\mu)
                        \right) 
                + C.
\label{Delta chi development}
\end{split}
\end{equation}
In the modified gravity field, the energy needed to escape the 
planet can be defined by the energy needed to reach the Roche lobe. 
Along the star-planet axis, the Roche lobe is located at a distance $r=r_{\rm Roche}$,
where $r_{\rm Roche}$ follows:
\begin{equation}
\frac{\partial \chi'}{\partial r}(r_{\rm Roche})=0.
\label{F=0}
\end{equation}
Again assuming that $r \ll  a_p$, and using Eq.~\ref{Eq chi'} and~\ref{F=0} we find 
\begin{equation}
\mu=3\left(\frac{r_{\rm Roche}}{a_p}\right)^3 + O\left(\frac{r_{\rm Roche}}{a_p}\right)^4,
\label{Eq:rRoche_mu}
\end{equation}
from which we derive
\begin{equation}
\frac{r_{\rm Roche}}{a_p}
\approx \left( \frac{\mu}{3}\right)^{1/3}.
\label{Eq:rRoche}
\end{equation}

Finally, the new potential energy of a unit mass of the atmospheric gas of a planet, 
$dE'_{p({\rm atm})}/dm$, defined by the opposite of the energy needed to
escape the planet gravity, is now the energy to reach the Roche lobe:
\begin{equation}
\begin{split}
dE'_{p({\rm atm})}/dm  \equiv & \chi'(R_p)-\chi'(r_{\rm Roche}) \\
  = &\left[\chi(R_p)-\chi(r_{\rm Roche})\right] \\
    & + \left[\Delta\chi(R_p)-\Delta\chi(r_{\rm Roche})\right].
\end{split}
\label{Eq App dE'}
\end{equation}
In the case $\mu \ll 1$ and $R_p\ll a_p$,
using Eq.~\ref{Eq chi} and~\ref{Eq:rRoche_mu} 
we derive for the first term of Eq.~\ref{Eq App dE'}:
\begin{equation}
\begin{split}
&\frac{\chi(R_p)
-\chi(r_{\rm Roche})}{G(M_p+M_*)}=
\\
&
     -\frac{\mu}{R_p}
     +\frac{3}{a_p}\left(\frac{r_{\rm Roche}}{a_p}\right)^2
     +\frac{1}{a_p}O\left(\frac{r_{\rm Roche}}{a_p}\right)^3.
\label{Eq dEp' 1st term}
\end{split}
\end{equation}
For the second term of Eq.~\ref{Eq App dE'}, using Eq.~\ref{Delta chi development} 
we derive:
\begin{equation}
\begin{split}
&
\frac{\Delta\chi(R_p)
-\Delta\chi(r_{\rm Roche})}{ G(M_p+M_*) }= 
\\
&    
 \frac{3}{2a_p}  \left(\frac{r_{\rm Roche}}{a_p}\right)^2 +
\\
& 
   \frac{1}{a_p}\left(O\left(\frac{R_p}{a_p}\right)^2
                               +O\left(\frac{r_{\rm Roche}}{a_p}\right)^3
                               +O\left(\mu\right)\right).
\end{split}
\label{Eq dEp' 2nd term}
\end{equation}
Here we implicitly assumed $(r_{\rm Roche}/R_p)^2\gg 1$, which is valid as the
smallest values in extreme cases of very-hot-Jupiters are always larger than 10.

As a result, combining Eq.~\ref{App Eq Ep atm}, \ref{Eq App dE'}, \ref{Eq dEp' 1st term} 
and~\ref{Eq dEp' 2nd term} we find 
\begin{equation}
\begin{split}
dE'_{p({\rm atm})}/dm \approx 
\ & \ 
dE_{p({\rm atm})}/dm 
\\
&
 + \frac{9G(M_p+M_*)}{2a_p}\left(\frac{r_{\rm Roche}}{a_p}\right)^2.
\label{App.Eq.dE'p.1}
\end{split}
\end{equation}
If we define $\Delta_{\rm tidal} dE_{p({\rm atm})}/dm$ by
\begin{equation}
dE'_{p({\rm atm})}/dm \equiv dE_{p({\rm atm})}/dm 
+ \Delta_{\rm tidal} dE_{p({\rm atm})}/dm,
\end{equation}
using Eq.~\ref{Eq:rRoche} and~\ref{App.Eq.dE'p.1}, we find 
\begin{equation}
\Delta_{\rm tidal} dE_{p({\rm atm})}/dm \approx
\frac{3^{4/3}G (M_p+M_*) \mu^{2/3}}{2a_p} .
\end{equation}
That is 
\begin{equation}
\Delta_{\rm tidal} dE_{p({\rm atm})}/dm \approx
\frac{3^{4/3}}{2} \frac{ G  M_{*}^{1/3} M_{p}^{2/3}}  { a_p }.
\end{equation}
This result is obtained using the potential description in the 
direction of the star on the star-planet axis, but it is valid
in all direction because the potential is the same on the whole
Roche lobe surface which is defined as an iso-potential.
Therefore the energy needed to reach the Roche lobe and to escape 
the planet is the same in all directions.

\section{The radial velocity iso-curves in the energy diagram}
\label{App:rad vel}

Most of the known planets have been detected with 
the radial velocity method which is not sensitive
to low mass planets with large orbital distances.
Therefore, if the detected planets are plotted in the energy diagram 
(Fig.~\ref{Ep_planet_vs_Euv}), the absence of planets 
with small absolute value of the potential energy (low mass)
receiving a low EUV energy (large orbital distance) is due to this
bias of the radial velocity method. 
The amplitude of the radial velocity of a planet-hosting star,
$V_{\rm rad}$, is given by
\begin{equation}
V_{\rm rad}^2=\frac{G M_p^2}{a_p M_*},
\end{equation}
where $G$ is the gravitational constant, $a_p$ the planet
orbital distance, $M_p$ and $M_*$ are 
the mass of the planet and the star, respectively.
Using Eq.~\ref{E_p}, and the fact that 
\begin{equation}
\frac{dE_{\rm EUV}/dt}{P_{*, {\rm EUV}}}=\frac{R_p^2}{4 a_p^2},
\end{equation}
where $P_{*, {\rm EUV}}$ is the total power emitted in EUV by the parent star,
we derive
\begin{equation}
E_p\approx-\frac{3}{8}M_*V_{\rm rad}^2
  \sqrt{\left(\frac{dE_{\rm EUV}/dt}{P_{*, {\rm EUV}}}\right)^{-1}}.
\end{equation}
This equation does not depend on the planetary radius, 
mass and orbital distance. 
Therefore, for a given stellar type, corresponding 
to a given stellar mass and total power emitted in the EUV,
the radial velocity iso-curves
must follow a straight line in the log-log energy diagram
of $E_p$ as a function of $dE_{\rm EUV}/dt$.

Most of the instruments 
have a detection threshold at about 10\,m\,s$^{-1}$. 
Instruments with larger sensitivity (like the {\sl Harps} spectrograph) 
are recent and 
allowed the discovery of low mass planets only with short orbital
periods. This explains why the detected planets below the limit
at 10\,m\,s$^{-1}$ are in the left part of the energy diagram
plotted in Fig.~\ref{Ep_planet_vs_Euv}.

\end{document}